\documentclass{PoS}
\usepackage{epsfig}

\newcommand{\AmS}{{\protect\the\textfont2
    A\kern-.1667em\lower.5ex\hbox{M}\kern-.125emS}}
\def\si{{}^1\kern-.14em S_0}
\def\siii{{}^3\kern-.14em S_1}
\def\piii{{}^3\kern-.14em P_1}
\def\diii{{}^3\kern-.14em D_1}
\def\lone{l_1}

\def\loneA{l_{1,A}}
\newcommand{\gsim}{\raisebox{-0.7ex}{$\stackrel{\textstyle >}{\sim}$ }}

\PoS{PoS(LAT2005)020}
\title{Nuclear Physics and Lattice QCD}
\ShortTitle{Nuclear Physics and Lattice QCD}
\author{\speaker{Martin J. Savage}\thanks{I would like to thank Daniel Arndt, Silas Beane, Paulo
  Bedaque, William Detmold, Kostas Orginos, Elisabetta Pallante and Assumpta Parre{\~n}o who
  are collaborators in this work.}\\\
                                                                           
        University of Washington\\
                                                                           
        E-mail: \email{savage@phys.washington.edu}}
       \abstract{Lattice QCD is progressing toward being able to 
impact our understanding of nuclei and nuclear processes.
I discuss areas of nuclear physics that are becoming possible to explore with 
lattice QCD, the techniques that are currently available and the status
of numerical explorations.
}
\FullConference{XXIIIrd International Symposium on Lattice Field
Theory\\
                                                                           
                 25-30 July 2005\\
                                                                           
                 Trinity College, Dublin, Ireland}
\begin{document}
\section{Introduction}

Nuclear physics is a well-established and mature field of study.  
A large number of experiments have been performed during the past decades that
have led to a great understanding of nuclei and nuclear processes.
We are currently moving into an exciting time for the field, one in which,
for the first time, properties of nuclei will be rigorously determined directly from QCD.
This will first occur through numerical calculations of quantities using lattice QCD,
matching onto effective field theories (EFTs) that have been developed during the
last decade or so, and using the EFT's to compute the
properties of the light nuclei. 
The first lattice
QCD calculation of the deuteron will be a milestone for nuclear physics.
It is this scenario that has motivated a  number of us to leave the relative
comfort of our ''analytic'' offices, in which we developed the
EFT's and fretted over the ever increasing number of
counterterms, and to attempt calculations of multi-nucleon systems with lattice QCD.

\section{Motivations}

There are a number of compelling reasons to use lattice QCD to perform nuclear
physics calculations.  
While it is important to explore and understand the building-blocks of
nuclei, the proton and neutron,  a greater challenge lies in 
understanding the structure and interactions of nuclei.

\subsection{The Emergence of Nuclear Physics from QCD}

One of the deep issues in our field,
one for which we currently have little or no
understanding, is how nuclear physics emerges from QCD.
The Lagrangian describing  QCD is simple: 
there is a 
gauge-covariant kinetic term for the quarks,
a mass-term for the quarks and the Yang-Mills term describing the glue.
There are a small number of fundamental constants 
that appear in this Lagrangian: $\Lambda_{\rm QCD}$ and the
six quark masses.
This Lagrangian and these seven constants (along with the electroweak
interactions) give rise to all of nuclear physics.
We wish to establish a rigorous pathway from QCD to nuclei.  
As mentioned previously, this will
be accomplished by performing lattice QCD calculations of some quantities,
determining counterterms in the appropriate low-energy EFT from these
calculations, 
and then using the EFT to compute quantities of interest in relatively simple systems.  
This EFT will be subsequently matched onto the many-body machinery that has been
developed and used by nuclear physicists, which will then be used to compute
the properties of more complex systems~\footnote{This may involve a lattice
  calculation of the effective field theory itself, 
e.g. Ref.~\cite{Muller:1998rc,Muller:1999cp,Abe:2003fz,Lee:2004qd}, 
  without reference to QCD.}.

\begin{figure}[!ht]
\centerline{\epsfig{file=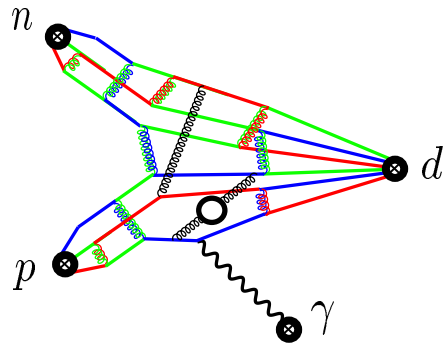, width=0.4\textwidth}}
\caption{A cartoon of the QCD calculation of the simplest nuclear
radiative capture process  $np\rightarrow d\gamma$.}
\label{fig:npqcd}
\end{figure}
As a simple example, consider the classic nuclear process 
$n p\rightarrow d \gamma$.
This is a process in which a low-energy neutron is captured by a
proton
to form a deuteron, emitting M1 radiation in the process.  
As the energy
increases from threshold the E1 amplitude rapidly increases and soon dominates the M1 amplitude.
A cartoon of the process at the quark-diagram level is shown in fig.~\ref{fig:npqcd}.
The Maiani-Testa theorem~\cite{Maiani:ca} prevents direct ``computation of this
process'' on the lattice.
Instead, one must compute other quantities on the lattice and use these
quantities to obtain  the matrix elements of interest.  This will be
accomplished by matching a lattice calculation to an EFT calculation.
For the process at hand, one can either use the pionless theory, in which only
nucleons and photons are dynamical degrees of freedom, or the pionful theory,
in which nucleons, pions and photons are dynamical, see Ref.~\cite{Beane:2000fx}.
The pionless theory is very useful for processes at momentum much below the
pion mass, and this theory has been explored out to several orders in the
power-counting.
The pionful theory is somewhat less developed, but if one is
interested in the quark-mass dependence, as will be the case for the initial
calculations performed at quark-masses greater than those of nature, the
pionful theory will be used.
In either theory, there are contributions that are unrelated to the NN
scattering amplitude, described by gauge-invariant operators involving nucleons
and the electromagnetic field strength tensor.
The coefficients of the operator(s) will be determined by
matching to a lattice calculation.  For this process ($np\rightarrow d\gamma$)
there are experimental
constraints that provide a very precise determination of these counterterms,
and a lattice calculation will provide (other than the quark mass dependence)
nothing more than a check of the method.

Ultimately, moving toward more complex systems one will need to match these
EFTs, the zero, one and higher nucleon EFTs to the many-body techniques that
have recently been developed.  The nuclear shell-model continues to move toward
a rigorous underpinning based on the renormalization group.
In principle,
one starts with a very large model space, as defined by the number of
levels of the harmonic oscillator (HO) (for instance) and matches chiral perturbation
theory and NN EFT's to observables in this model space.  One then
systematically reduces the model space, by integrating out the high-levels of
the HO. This is exactly analogous to the renormalization group methods we are
used to dealing with in particle physics, except this is now constructed for
bound states.  As states are integrated out, the Hamiltonian and operators are
renormalized.  Eventually one is left with a small, low-energy space --the
shell-model space--and a large excluded space.   The operators and Hamiltonian
evolve in such a way that the energy-levels and matrix elements in the
shell-model space reproduce those computed in the full-space, see Ref.~\cite{Haxton:1999vg}.

\subsection{Calculations of Processes where Experiments are not Possible}

Lattice QCD will have significant impact on nuclear physics by computing
quantities for which there is little or no data.  One such example of this is
to perform calculations that impact our understanding of dense nuclear matter.
I do not mean asymptotically dense, but at
the densities that arise in the collapse of stars,
that are a few times nuclear matter density.
A second example  is the calculation of matrix elements of operators
that probe physics beyond the standard model, such as $\beta\beta$-decay.

\subsubsection{Supernova, Neutron Stars and Black Holes}

The nature of the remnant of supernova 1987A
is still undetermined.  It is unknown if the remnant is a neutron star or a
black hole (and it is unknown if it underwent spherical collapse or aspherical collapse).
At the heart of the matter is nuclear physics, and in particular the
compressibility of nuclear matter at densities that are a few times that of
ordinary nuclear matter.  To resolve this issue, one needs to know the
composition of nuclear
matter in this regime, and this is where lattice QCD can contribute.

As one compresses nuclear matter composed of neutrons, protons and electrons, 
the Fermi-energy of the electronic
component increases rapidly compared to that of the nucleons.
At some point it becomes energetically favorable for the system to increase
the density of neutrons via the weak interaction,
rather than to further increase the density of electrons and protons, despite the 
proton-neutron mass-splitting.
Hence, the name neutron star.  
Thus it is the neutrons and their interactions that
primarily dictate the compressibility of nuclear
matter at densities higher than nuclear matter densities.
This discussion ignores the possibility that other hadronic components can
become energetically favored.  
In 1986 Kaplan and Nelson~\cite{Kaplan:1986yq,Nelson:1987dg} 
pointed out that the $K^- n$ interaction should
significantly reduce the mass of the $K^-$ in neutron matter, and as stressed
by Bethe and Brown~\cite{Brown:1993jz}, 
if the $K^-$ mass falls below the electron Fermi
energy, the formation of a condensate of $K^-$'s, $\langle K^-\rangle\ne 0$
will occur.  
Nuclear matter would then be composed of neutrons, protons and a kaon condensate
which has  a softer equation of state than pure neutron
matter.  The $nK^-$ scattering amplitude is poorly known, and thus there is
considerable uncertainty in the density dependence of the $K^-$ mass.
This ambiguity can be resolved by a careful lattice QCD study.

A competing process arises from the density dependence
of the $\Sigma^-$ mass.  If the mass of the $\Sigma^-$ drops below the electron
chemical potential, then nuclear matter will be composed of neutrons, protons
and $\Sigma^-$'s.  
For obvious reasons, this system will have a softer
equation of state than pure neutron matter.
The $n\Sigma^-$ scattering amplitude is also poorly known, only accessible through
the properties of hypernuclei, and so one can presently 
only speculate as to the existence of such matter.

Lattice QCD can have significant impact on determining the equation of state of
hadronic matter at high densities.  The two processes discussed above,
kaon-condensation and hyperonic matter can be refined by a calculation of the
neutron-kaon and neutron-hyperon scattering amplitudes.

\subsubsection{$\beta\beta$-Decay}

The last decade has seen remarkable progress in our understanding of the
neutrino sector.  The  measurements by SuperK and SNO demonstrating that neutrinos
have non-zero mass, and that the mass eigenstates are not the flavor eigenstates,
confirmed what had been speculated for many years based upon the results of 
the Davies chlorine experiment.  It remains to be determined if the neutrinos
are Dirac or Majorana.  In the later case, lepton number is not conserved, as
the neutrino is its own anti-particle.  Observation of a lepton number
violating process, such as neutrinoless $\beta\beta$-decay, would be yet another sensational
discovery in the neutrino sector.

Nature has provided a limited number of nuclei that undergo $\beta\beta$-decay,
and 
the constraints on building detectors to observe $\beta\beta$-decay allow for
only a handful of elements to be considered.  Perhaps the most widely known is
Germanium, where one looks for a peak in the $e^-e^-$ invariant mass spectrum
resulting from $^{76} Ge\rightarrow ^{76} Se\  e^- e^-$, sitting on top of the
background from the lepton number conserving process
$^{76}Ge \rightarrow ^{76}Se\  e^- \overline{\nu}_e  e^- \overline{\nu}_e$
as a signal of lepton-number violation.
As the simplest baryonic process that can contribute to this process is of the
form $p p\rightarrow n n e^- e^-$, and higher body operators are expected to be
suppressed, it is clear that a rigorous calculation of the rate of such decay 
is not simple.  There are complications at the many-body level, 
where the calculation of the nuclear rate from the two-nucleon operator (local
or non-local) is notoriously difficult.
Further, for a massive neutrino, or where the process is dominated
at short-distances, matching from QCD onto the nuclear EFT suffers from issues 
similar to those
that arise in the calculation of
nonleptonic weak matrix elements~\cite{Savage:1998yh,Prezeau:2003xn}.
Lattice QCD is the only way to rigorously explore these matrix elements.

\subsection{The Dependence of Nuclei and Nuclear Processes on the fundamental
  constants of Nature}

Before we can be content with our understanding of nuclei and nuclear processes
we must know and understand how they depend upon the fundamental constants of
nature.
The recent suggestion that the electromagnetic coupling is time dependent,
based upon the splitting between atomic absorption-lines in gas clouds at different
red-shifts (along the line of sight to quasars)~\cite{Murphy:2000pz}, 
renewed interest in such a possibility.
I look forward to an independent verification of this exciting result.

If we consider a nuclear process that is of central importance to the
development of life, 
e.g. $3\alpha\rightarrow ^{12} C$, we should be able to fully
understand how the rate for this process, and the rate for the production of
carbon in stars depends upon the mass of the light-quarks, the scale of
the strong interaction and the value of the electromagnetic coupling constant.
At this point in time we have very little control over any aspect of this
calculation.
One of the reasons for this is that this process is highly fine-tuned, 
but also we have only  a limited
knowledge about how nuclear physics depends upon the quark masses (even in the
absence of fine-tunings).
Clearly it is not going to be easy to map out the spectrum of $^{12}C$ as a
function of the quark masses, however, this is a worthy goal for nuclear physics.

\subsection{Fine-Tunings in Nuclear Physics}

During the last few years the role of fine-tunings in physics has become a
central focus of particle physics.  
The question of why the cosmological constant, $\Lambda\sim 10^{-47}~{\rm GeV}^4$,
is so small and yet nonzero has sparked great interest.
However, fine-tunings also play a  central role in the strong interactions.

The nucleon-nucleon potential can be dissected into roughly three length
scales.
The long-distance part is well described by one pion exchange (OPE).
Performing a   phase shift analysis of NN scattering data while
treating the mass of the pion as a free parameter, along with its couplings,
recovers the physical pion mass and its axial coupling.
What is more impressive is that a recent analysis of NN
scattering~\cite{Rentmeester:2003mf}
has shown that the two-pion exchange contribution computed in chiral
perturbation theory provides a better fit than the traditional
``$\sigma$-meson'' exchange.  
What's really impressive is that  the values of the local counterterms that
enter at this order are
consistent with those extracted from the single-nucleon sector.
Therefore both the long-distance and
intermediate-distance component of the NN-potential is well described by chiral
perturbation theory.
While many ``old-school'' NN potentials have a one meson exchange
short-distance repulsion, the modern potentials 
have a short-distance component  that is optimized in shape and strength,
consistent with EFT power-counting, to best fit the data.

The truly fascinating aspect of this, is that while the NN potential has three
quite distinct distance scales, the s-wave wavefunction resulting from solving the
Schr{\"o}dinger equation for the two-nucleon system 
is approximately horizontal at the edge of the potential, i.e.
the scattering lengths are unnaturally large, $\sim -24~{\rm fm}$ in the $^1 S
_0$-channel and  $\sim +5.4 ~{\rm fm}$ in the $^3 S _1$-channel (the channel
containing the deuteron, which is bound by $\sim 2.2~{\rm MeV}$).
There is clearly a fine-tuning among the regions of the potential in order
for this to be true.  Clearly, this is all controlled by the parameters of QCD,
but in nature these parameters have conspired to place us very near an infrared
fixed-point in the hadronic sector (if the scattering lengths were infinite the
system would exhibit scale-invariance and would be at an infrared,
unstable fixed-point)~\footnote{The fact that both channels, and not just one, 
are close to the fixed point is understood in the large-$N_c$ limit of 
QCD~\cite{Kaplan:1995yg} leaving us to understand why the system as a whole is 
near the fixed-point}.
Proximity to this fixed point means that the EFT used to describe such systems
does not follow the usual power-counting developed for systems that have
natural scattering lengths, where it is simply a matter of counting engineering
dimensions.  
The four-nucleon operators contribute at the same order in the counting as the
chiral limit of OPE, while deviations from the chiral limit are suppressed~\cite{Beane:2001bc}.
In order to have states near threshold, the coefficients of the four-nucleon
operators are fine-tuned to achieve the delicate cancellation between kinetic and
potential energy that exists.

The $3\alpha\rightarrow ^{12}C$ process that was discussed previously, is also
finely-tuned.
In order for this process to proceed with any appreciable rate there are two
important fine-tunings that have occurred.  First, the ground state of $^8 Be$
is barely unbound, and in the stellar interior one has $2\alpha\rightarrow ^8
Be$, and the $^8 Be$ state lives for an unnaturally long period of time,
enough time for a third $\alpha$ to interact to form $^{12}C$ through radiative
capture.  However, the rate would still be insufficient except for the fact
that there is a level in $^{12}C$ at precisely the right energy to
significantly enhance the capture rate.
It is quite easy for me to believe that we can only understand this set of
fine-tunings via the anthropic principle, as without carbon we would not be
having this discussion.  However, I have been unable to reconcile these
fine-tunings with those in the NN-sector, as I have no reason to believe that
the form of the NN interaction required to greatly enhance the triple-alpha
process would also put us close to a fixed-point in the NN-sector.  There has
been no work to try to reconcile these features, and it is something that
should be understood.

\section{The Quark-Mass Dependence of Nucleon-Nucleon Scattering}

\begin{figure}[!ht]
\centerline{\epsfig{file=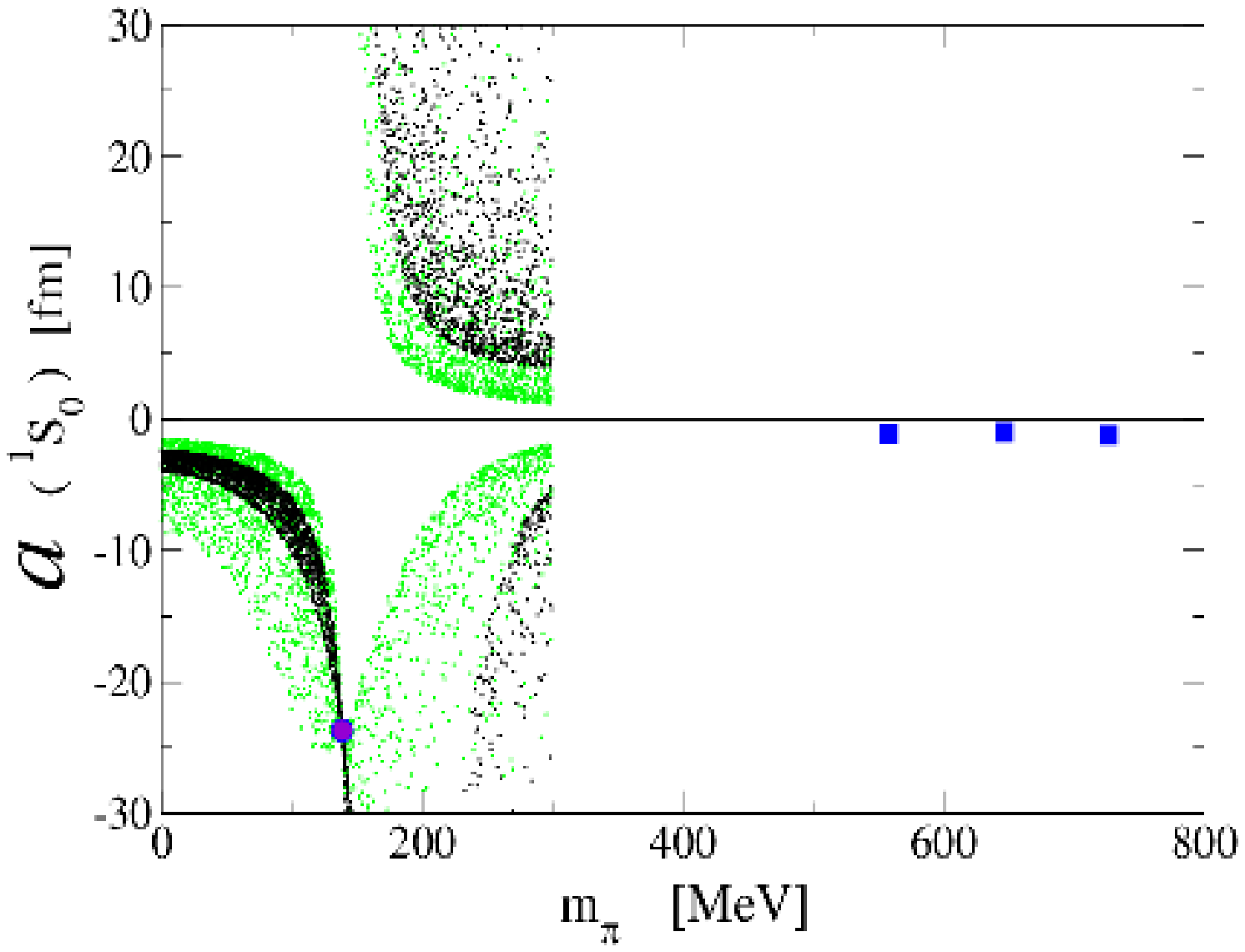, angle=0,width=0.5\textwidth}
\epsfig{file=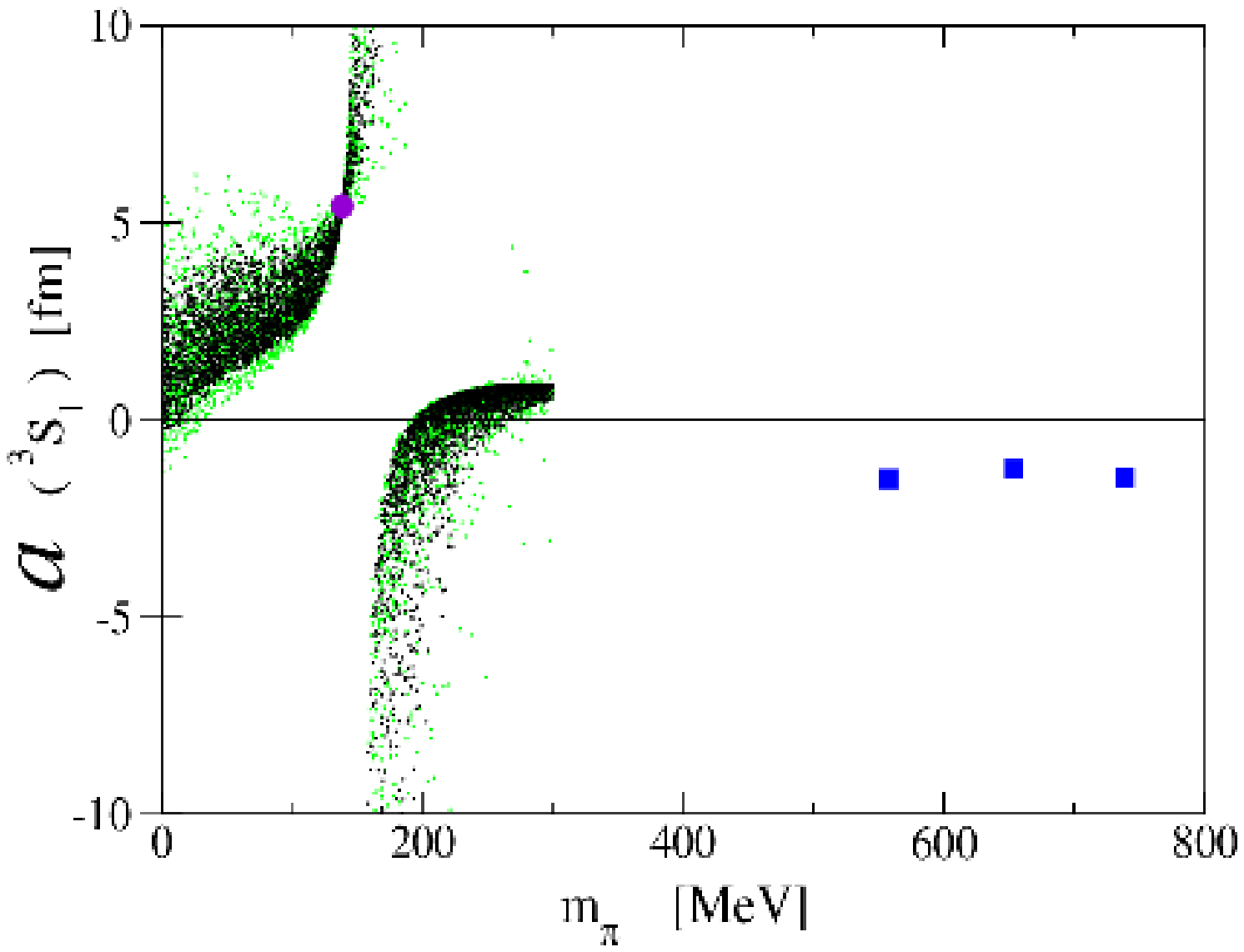,  angle=0,width=0.5\textwidth}}
\caption{The NN scattering lengths in the s-wave channels as  a function of the
  pion mass at NLO in the NN EFT.  The two regions defined by the points (green
  and black) 
  correspond to two different choices of the (sets of) coefficients of the
  $m_q$-dependent four-nucleon operators~\protect\cite{Beane:2002xf}.
The three (blue) solid squares are the scattering lengths resulting from a
quenched lattice QCD calculation~\protect\cite{Fukugita:1994ve}.
}
\label{fig:amq}
\end{figure}
The development of EFTs for nuclear physics during the last 15 years since
Weinberg's pioneering papers on the subject, have allowed us 
to rigorously explore the quark-mass dependence of simple nuclear systems.
We have learned that simple analytic dependences on the quark masses are
incorrect and the actual quark mass dependence is quite complex.
This is compounded by the fine-tunings discussed in the previous section.

During the last couple of years, the quark-mass dependence of the two-nucleon
sector has been explored~\cite{Beane:2002vs,Beane:2002xf,Epelbaum:2002gb}.
The scattering lengths as a function of the pion mass are shown in fig.~\ref{fig:amq}.
At the first non-trivial order, in addition to the quark mass dependence 
from OPE, there are also contributions from four-nucleon operators that
have a single insertion of the quark mass matrix with coefficients, the $D_2$'s,  that are
currently unknown.  To determine the reasonable ranges of scattering length at
a given quark mass dimensional analysis has been used to constrain the
coefficients of these operators.  
Fig~\ref{fig:amq} suggests that
it is likely that the di-neutron remains unbound as one moves toward the chiral
limit, while it could be bound or unbound as the mass is increased.  In the
deuteron channel, the deuteron binding energy moves toward its natural value
for decreasing quark mass, but again, it could be bound or unbound at higher
masses.
In fig.~\ref{fig:amq} I have shown the only lattice QCD calculation of this 
system~\cite{Fukugita:1994ve}.
This pioneering lattice 
calculation is quenched and at large pion masses, outside the range of
the EFT's that currently exist, and therefore, unfortunately, 
does not shed light on the
question at hand.

The program to undertake is to perform a lattice QCD calculation of the
scattering lengths (or of the deuteron) at the larger masses that are currently
accessible, but inside the range of validity of the EFT, match to the $D_2$'s,
and use the EFT to make the chiral extrapolation to map out the entire region
of interest.

\section{Two Hadrons on the Lattice}

\begin{figure}[!ht]
\centerline{\epsfig{file=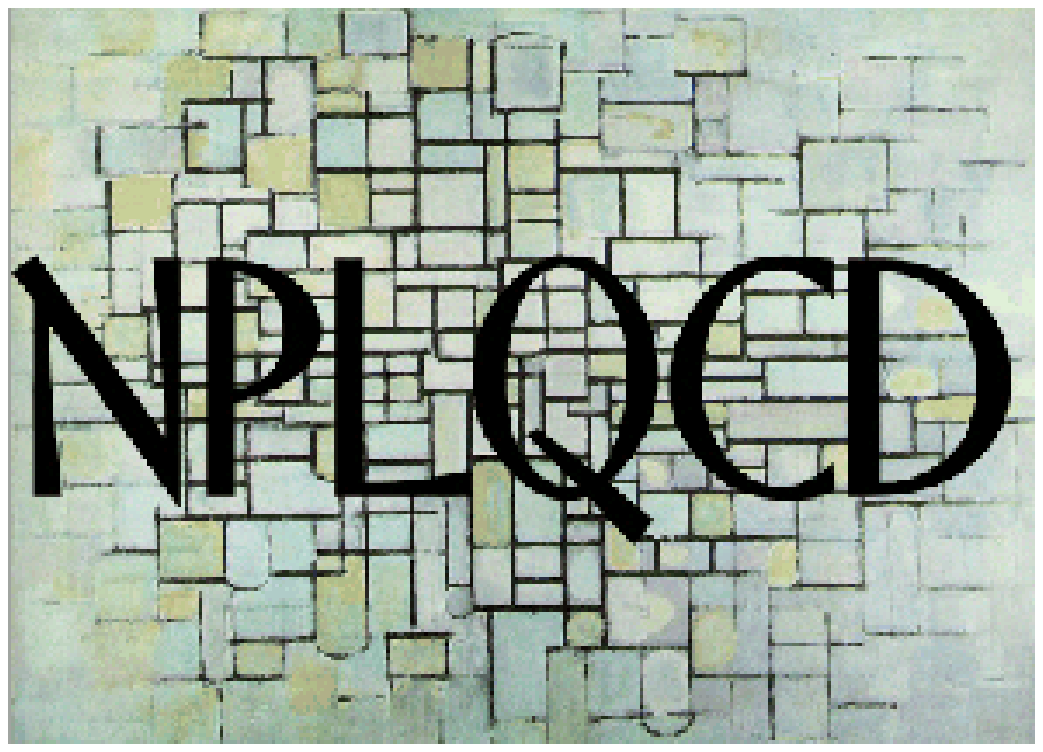, angle=0,width=0.4\textwidth}\qquad\qquad
\epsfig{file=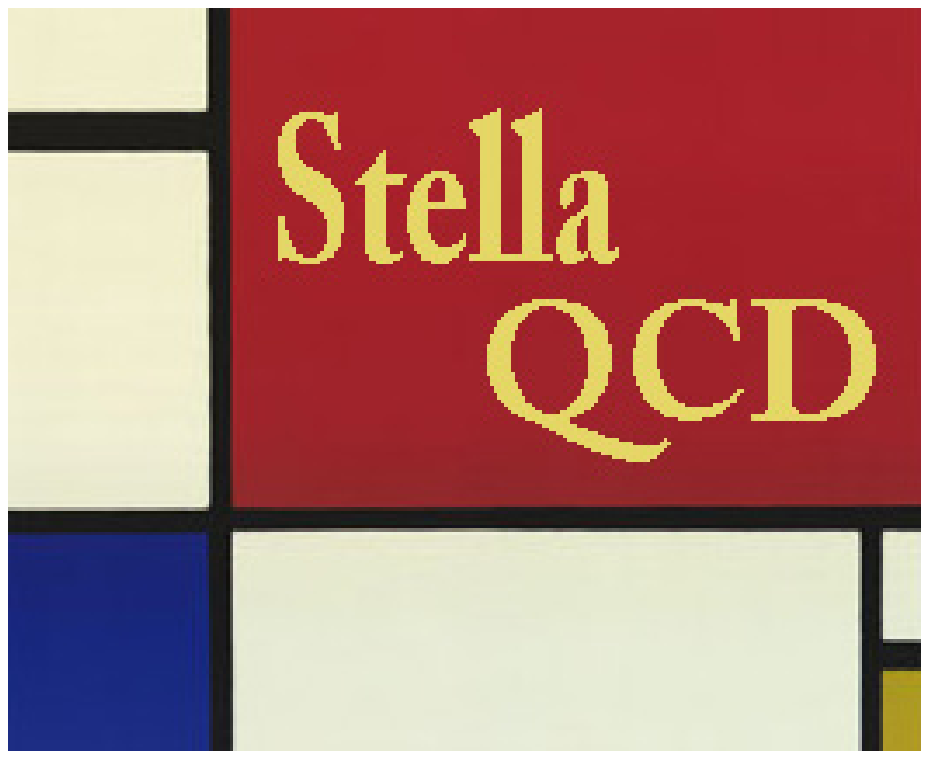,  angle=0,width=0.4\textwidth}}
\caption{The logos of NPLQCD and StellaQCD.
}
\label{fig:logos}
\end{figure}

Motivated by the physics described in the previous sections two collaborations
have emerged: NPLQCD and 
StellaQCD, see fig.~\ref{fig:logos}
~\footnote{The participating institutions are 
University of Washington,
University of New Hampshire, Lawrence Berkeley Laboratories, College of
William and Mary, University of Groningen and University of Barcelona.
}.
The current focus of these two efforts is to extract the scattering lengths,
and more generally, the 
phase shifts associated with low-energy nucleon-nucleon, nucleon-hyperon and
hyperon-hyperon scattering.
%

\subsection{The Maiani-Testa Theorem (The End of the Innocence) and L{\"u}scher's Method}

The first hurdle that one encounters when considering the extraction of
a scattering amplitude from a lattice QCD calculation is the Maiani-Testa
theorem.  Essentially no-one outside of the conference attendees appreciates
the implications of this theorem, and typically the impression that most have
is that one will be able to calculate everything directly from lattice QCD.
Unfortunately, this is false.
The Maiani-Testa theorem states that one cannot extract two-body (and higher)
S-matrix elements from Euclidean space Green functions at infinite-volume except
at kinematic thresholds.  Obviously, this has huge implications for the
computation of nuclear processes, even the simple ones like $np\rightarrow
d\gamma$, with lattice QCD.

In the two-body sector, L{\"u}scher was able to arrive at a method to circumvent
this constraint~\cite{Luscher:1986pf,Luscher:1990ux}. 
Below inelastic threshold the volume dependence of the
two-particle energy levels is determined by the scattering amplitude.
Thus determining the energy-levels of two particles in a finite-volume allows
for a determination of the scattering amplitude (at the location of the energy-levels).
L{\"u}scher's formula relating the scattering amplitude to the location of
two-particle states at finite-volume is 
\begin{eqnarray}
p\cot\delta(p) \ =\ {1\over \pi L}\ {\bf S}\left(\,\left({Lp\over 2\pi}\right)^2\, \right)\ \ ,
\label{eq:energies}
\end{eqnarray}
where
\begin{eqnarray}
{\bf S}\left(\,{\eta}\, \right)\ \equiv \ \sum_{{\bf j}}^{\Lambda_j} 
{1\over |{\bf j}|^2-{\eta}}\ -\  {4 \pi \Lambda_j}
\ \ \  ,
\label{eq:Sdefined}
\end{eqnarray}
and it is understood that the UV cut-off, $\Lambda_j\rightarrow\infty$.
In particular limiting cases one can arrive at analytic expressions for both the
continuum states and bound states.
For $a/L\rightarrow 0$, the lowest energy state is located at (we are using the
nuclear physics definition of the scattering length)
\begin{eqnarray}
E_0 & = & + {4\pi a\over M L^3}\left[\ 1\ -\ c_1 {a\over L}\ 
+\ c_2 \left({a\over L}\right)^2\ +\ ...\right]
\ +\ {\cal O}(L^{-6})
\ \ \ ,
\label{eq:e0}
\end{eqnarray}
where the coefficients are $c_1=-2.837297$, $c_2=+6.375183$,
and the bound state, if one exists, is located at 
\begin{eqnarray}
E_{-1} & = & -{\gamma^2\over M}\left[\ 
1\ +\ {12\over \gamma L}\  {1\over 1-2\gamma (p\cot\delta)^\prime}\ 
e^{-\gamma L}\ +\ ...
\right]
\ \ \ ,
\label{eq:eb}
\end{eqnarray}
where $(p\cot\delta)^\prime={d\over dp^2}\ p\cot\delta$ evaluated at
$p^2=-\gamma^2$. The quantity $\gamma$ is the solution of
\begin{eqnarray}
\gamma\  +\  p\cot\delta |_{p^2=-\gamma^2} \ & = & 0
\ \ \ .
\label{eq:pctdg}
\end{eqnarray}
In the case of $a/L\rightarrow\infty$ (more precisely $L p\cot\delta \rightarrow
0$), of relevance to the NN system, one finds that
\begin{eqnarray}
\tilde E_0 & = & 
{4\pi^2\over M L^2}\left[\ d_1 \ +\   d_2\  L p\cot\delta_0\  + ...\
  \right]
\ \ \ ,
\label{eq:usE0}
\end{eqnarray}
where the coefficients are $d_1 = -0.095901$, $d_2 = +0.0253716$ and
where $p\cot\delta_0$ is evaluated at an energy $E={4\pi^2\over M L^2}\ d_1$.

\subsection{The Present and the Future for NN Scattering}

One of the exciting aspects of studying two-nucleons on the lattice is that
rigorous calculations can be performed today, at unphysical pion masses,
even if the systems have scattering lengths that are much larger than the
lattices. 
The reason for this is that it is the range of the interaction that is
relevant, and not the scattering length.  The range of the NN potential is
dictated by the mass of the pion, and while the physical pion mass requires
volumes larger than $\gsim 5~{\rm fm}$, a pion of mass $\sim 350~{\rm MeV}$
requires volumes larger than $\gsim 2.5~{\rm fm}$, which means that 
meaningful calculations can be performed on the available MILC
lattices.

\begin{figure}[!ht]
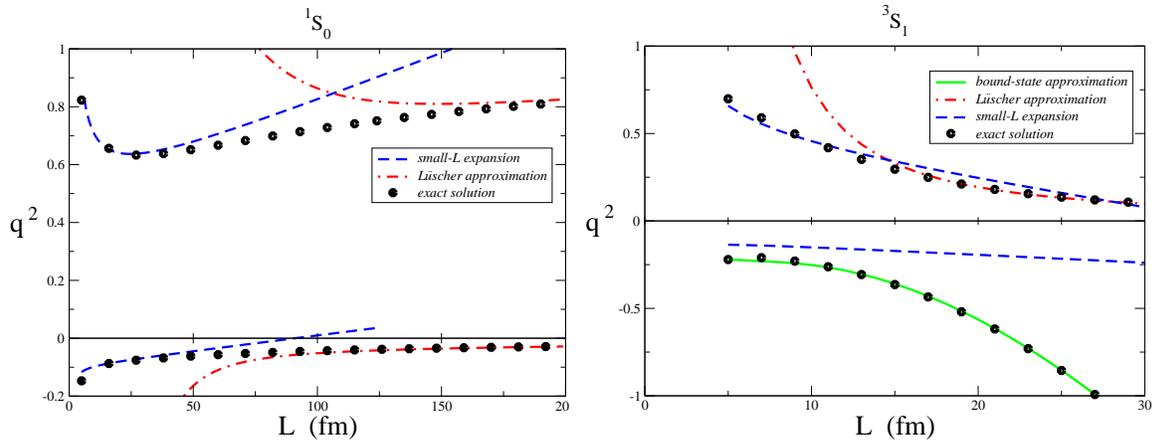

\centerline{\epsfig{file=SINGLET.eps, angle=0,width=0.5\textwidth}
\epsfig{file=TRIPLET.eps,  angle=0,width=0.5\textwidth}}
\caption{The position of energy levels of two nucleons in a cubically-symmetric
  finite volume of spatial dimension $L$ at the physical value of the pion
  mass. The y-axis is $q^2\sim E_n L^2$. The points correspond to exact
  solution of eq.~(\protect\ref{eq:energies}), 
while some of the curves correspond to the approximations
in eq.~(\protect\ref{eq:e0}), eq.~(\protect\ref{eq:eb}), and
eq.~(\protect\ref{eq:usE0}).
As the interactions between nucleons in the two channels are similar, the
spectra in small volumes are also similar.
Further, as the scattering lengths are large in both channels (a fine-tuning
between potential and kinetic energy), the 
spectra in large volumes are similar, except for the bound state in the 
channel with positive scattering length.
}
\label{fig:NNbox}
\end{figure}
The simplest tool for analyzing lattice results is the pionless EFT, which
is appropriate below the pion inelastic threshold.
Beane, Bedaque, Parre{\~n}o and I~\cite{Beane:2003da} 
explored what would be found in  a lattice QCD
calculation of the NN system at finite volume at the physical value of the pion
mass using the well known NN scattering amplitude.
The results are shown in fig.~\ref{fig:NNbox}.
The energy of levels associated with two nucleons in the 
$\si$ and the mixed $\siii-\diii$ channels were computed, and 
solutions to 
eq.~(\protect\ref{eq:energies})
(containing both the continuum states and bound
state) are shown by the solid circles in fig.~\ref{fig:NNbox}.
As expected the bound state solution asymptotes at large volume to the deuteron
binding energy, while the lowest continuing levels move toward vanishing energy.

To explore NN interactions in smaller volumes, one needs to solve the pionful
theory in a finite-volume with periodic boundary conditions (or whatever
boundary conditions one chooses, for instance, twisted boundary conditions as 
first suggested by Bedaque~\cite{Bedaque:2004kc} and explored further in 
Refs.~\cite{Sachrajda:2004mi,Bedaque:2004ax}).
This has not yet been performed, and given that the pionful NN EFT calculations
are performed to lower orders than the pionless calculations, this is going to
require some serious efforts by the NN EFT community.

\subsection{Electroweak Interactions}

Perhaps the simplest non-trivial process in nuclear physics involving
electroweak interactions is the radiative capture process $np\rightarrow
d\gamma$.
This is a classic nuclear physics process, and provides a clean demonstration
of meson-exchange-currents (MEC's) (electroweak interactions that are not
related by gauge symmetry to the NN scattering amplitude).
Lattice QCD is going to be able to compute this process by matching onto the
low-energy EFT (either pionless or pionful), determining the coefficients of
the gauge-invariant operators that are not related to the NN scattering
amplitude, and using the EFT to compute the amplitudes of interest.
Detmold and I explored what could be learned from lattice QCD calculations of
processes such as this using background-field techniques~\cite{Detmold:2004qn}.
We found an expression for the energy-levels of two nucleons
in a finite volume in the presence of a background magnetic field, and also a
background $Z^0$ field (for the weak-dissociation of the deuteron).
The background fields mix the $^1 S _0$ and the $^3 S _1$ states, and the
spectrum is somewhat complicated.
However, there is at least one level in the spectrum (and the
challenge is to find lattices for which it is the ground state or a low lying
excitation) that is sensitive to the coefficient of the gauge-invariant operator.
This sensitivity comes about because the deuteron and the near bound state are
finely-tuned, and placing the system in a finite volume disturbs
this fine-tuning.  The contribution of the local operator
can restore the fine-tuning in the presence of the
background field.  It is in this scenario that the
level becomes sensitive to the value of the coefficient.

\begin{figure}[!ht]
\centerline{\epsfig{file=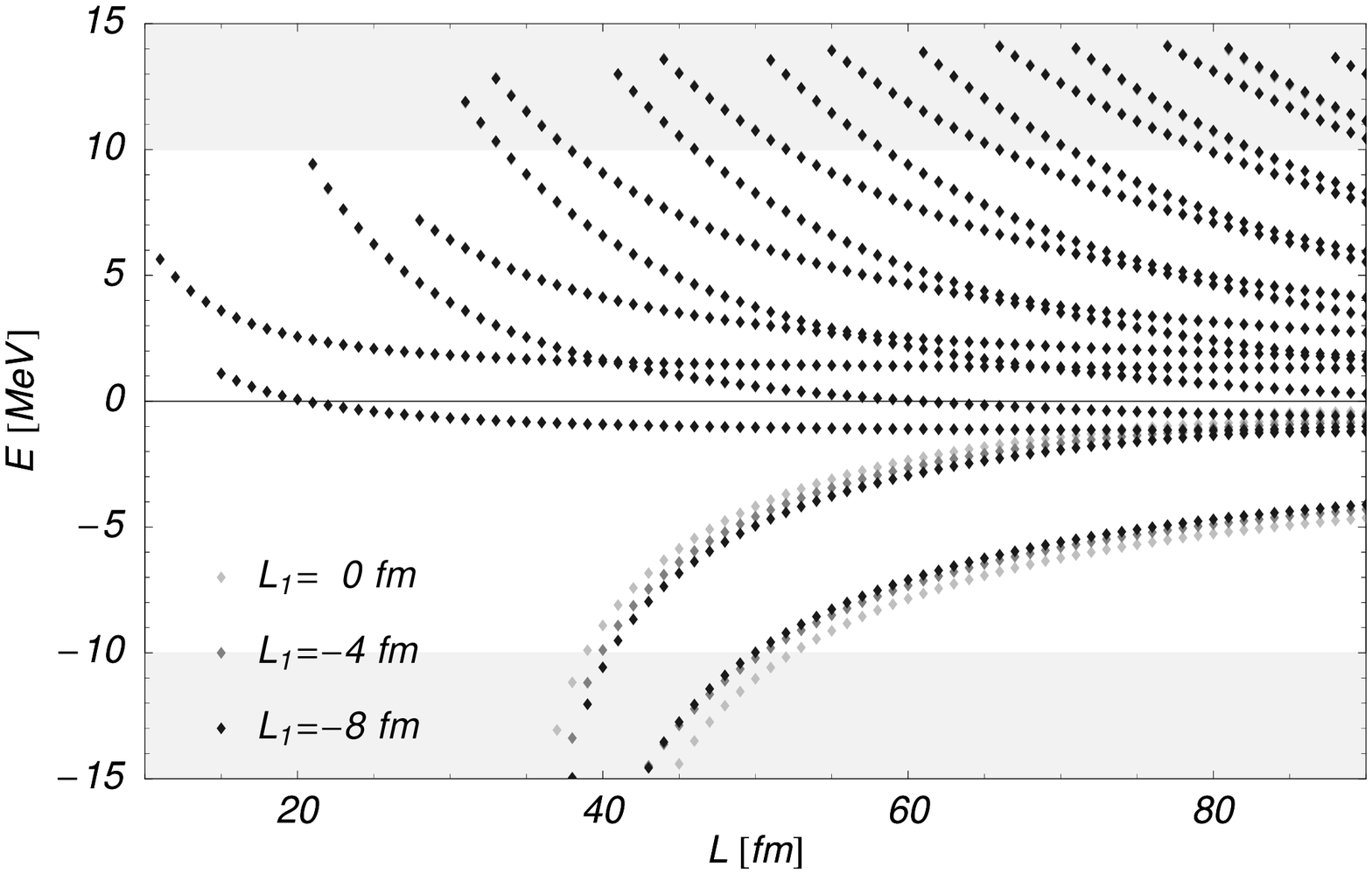, angle=0,width=0.5\textwidth}
\epsfig{file=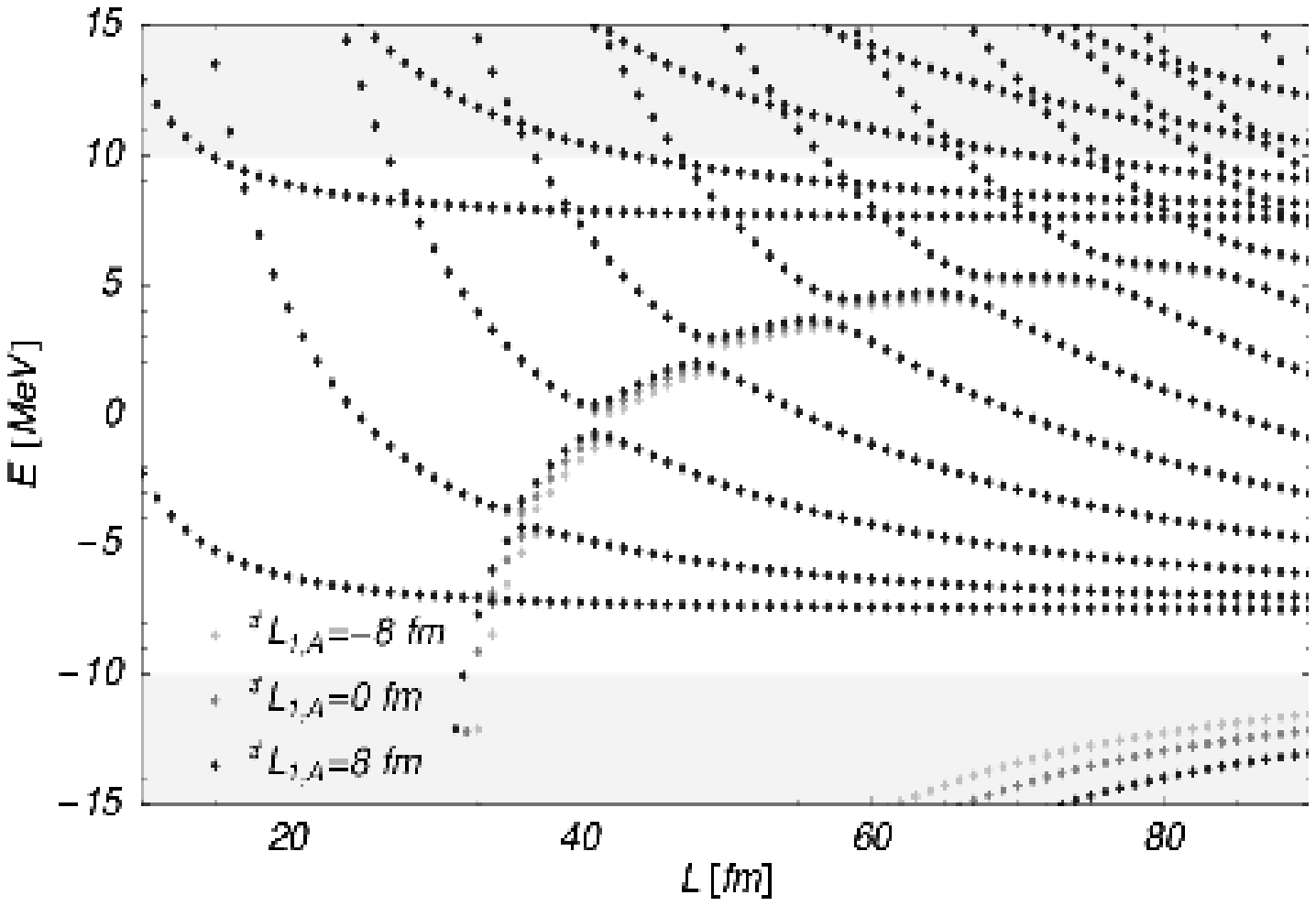,  angle=0,width=0.5\textwidth}}
\caption{NN energy levels in an asymmetric volume in the presence
  of a background magnetic field (left panel) and background $Z^0$-field (right
panel)~\protect\cite{Detmold:2004qn}.
There is at least one level that is sensitive to the value of the
gauge-invariant operator involving two nucleons and the background field.
}
\label{fig:bkgdNN}
\end{figure}

For  $np\rightarrow d\gamma$, the energy of two nucleons in a background field 
(in a cubically symmetric lattice are)
\begin{eqnarray}
\left[\ p\cot\delta_1 -{S_1 + S_2\over 2\pi L}  \right]
\left[\ p\cot\delta_3 -{S_1 + S_2\over 2\pi L}  \right]
  = 
\left[\ {e B_0 \lone \over 2} + {S_1 - S_2\over 2\pi L}  \right]^2 \ \  ,
\label{eq:m0solve}
\end{eqnarray}
where
\begin{eqnarray}
S_1 & = & S(\tilde p^2 + \tilde u_1^2)
\ \ ,\ \ 
S_2 \ = \ S(\tilde p^2 - \tilde u_1^2)
\ \ ,\ \ 
\tilde u_1^2 \ = \ {L^2\over 4\pi^2} \ e B_0 \ \kappa_1
\ \ \ ,
\label{eq:s1s2}
\end{eqnarray}
$\kappa_1=(\kappa_p-\kappa_n)/2$ and $p\cot\delta_1=-\frac{1}{a_1} +
\frac{r_1}{2}p^2 + \ldots$ is the $\si$ effective range expansion. The
$\tilde u_1$ contributions result from the one-nucleon interactions with
the background field, while the two-nucleon--background field interactions are described by
$\lone$.
It turns out to be desirable to work with asymmetric volumes to suppress the
Landau-level structure, and fig.~\ref{fig:bkgdNN} shows the levels for particular values
of the parameters.
For  $np\rightarrow d Z^{0*}$, the relation becomes 
\begin{eqnarray}
\left[\ p\cot\delta_1 -{S_1 + S_2\over 2\pi L} \right]
\left[\ p\cot\delta_3 -{S_1 + S_2\over 2\pi L}  \right]
\  = \
\left[ {gW\  \loneA \over 4} - {S_1 - S_2\over 2\pi L}  \right]^2
\ \ \ ,
\label{eq:m0solvewk}
\end{eqnarray}
where
\begin{eqnarray}
S_1 & = & S(\tilde p^2 + \tilde w_1^2)
\ \ ,\ \ 
S_2 \ = \ S(\tilde p^2 - \tilde w_1^2)
\ \ ,\ \ 
\tilde w_1^2 \ = \ - {L^2\over 4\pi^2} \ gW\ g_A\ M 
\ \ \ .
\label{eq:s1s2wk}
\end{eqnarray}
The energy-levels in an asymmetric volume are shown in fig.~\ref{fig:bkgdNN}.

\subsection{The $\Lambda_Q\Lambda_Q$ Potential}

\begin{figure}[!ht]
\centerline{\epsfig{file=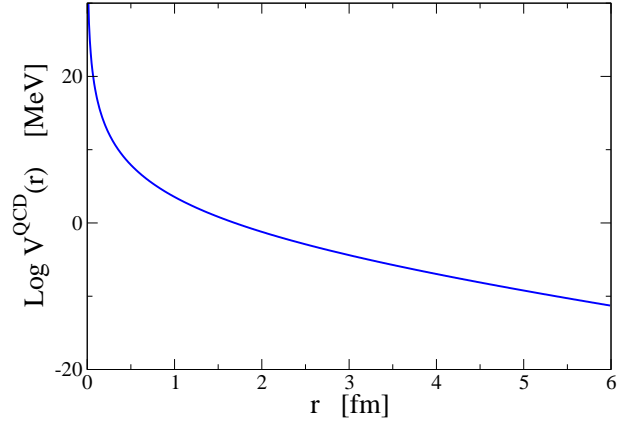, angle=270,width=0.6\textwidth}}
\caption{Logarithm of the potential between two $\Lambda_Q$'s.}
\label{fig:vrLQLQ}
\end{figure}

An observable that may impact our understanding of nuclear forces is the
potential between two $\Lambda_Q$ baryons in the heavy-quark limit.
The potential between to B-mesons in the heavy-quark limit has been explored
previously~\cite{Michael:1999nq,Pennanen:1999xi,Cook:2002am,Cook:2005ky},
both in quenched and dynamical calculations.
This interaction also resembles the nuclear force, including OPE.
For two $\Lambda_Q$ baryons,
isospin symmetry forbids contributions from OPE, and the leading long-distance
behavior arises from the exchange of two pions (TPE).
While TPE leads to a better description of the NN scattering phase-shift data
than heavy meson exchange, one would like to better understand this component
of the interaction.  The TPE contribution to the $\Lambda_Q$ potential is only
part of the TPE contribution to the NN potential, as the Weinberg-Tomazawa term
is absent, the axial matrix elements are different, and the spectrum of the
$\Lambda_Q-\Sigma_Q$ sector differs from the $N-\Delta$ sector.  However, one
will be able to directly explore what TPE ``looks like'' in this system, for
instance, if this part of TPE well-described by perturbation theory.
Arndt, Beane and I computed this potential~\cite{Arndt:2003vx} in 
$\chi$PT and also
in partially-quenched $\chi$PT.
As expected it ``falls like a rock'', as the relevant mass-scale is $2 m_\pi$, a
plot of which can be found in fig.~\ref{fig:vrLQLQ}.
%

\section{Nonleptonic Hyperon Decays}

As the nonleptonic decays $\Lambda\rightarrow p\pi^-$,
$\Sigma^-\rightarrow n\pi^-$, $\Sigma^+\rightarrow n\pi^+$,
$\Xi^-\rightarrow\Lambda\pi^-$ and their isospin partners have been
well studied for an extensive period of time, the experimental data
and theoretical efforts to understand the data can be found in many
textbooks.  Recently we emphasized that this is a problem worthy of exploration
with lattice QCD and examined the amplitudes using  SU(2)
symmetry~\cite{Beane:2003yx}. 

While the axial matrix elements $g_A\sim1.27$ and
$g_{\Lambda\Sigma}\sim 0.60$ are well-known experimentally, only an upper
limit currently exists for $\Sigma^-\rightarrow\Sigma^0
e\overline{\nu}$, and thus there is no direct measurement of
the coupling $g_{\Sigma\Sigma}$, which contributes to the $\Sigma$-decay p-wave
amplitudes. Orginos has been awarded time by SciDAC to compute this matrix element,
and the other octet-baryon axial matrix elements.

Inserting numerical values for the couplings and masses into the $S-$wave and
$P-$wave amplitudes gives the
numerical values for the nonleptonic amplitudes shown in table~\ref{table:su2}.
\begin{table}[ht]
\caption{Weak Amplitudes in $SU(2)$ $\chi$PT at LO ($g_{\Sigma\Sigma}=0.30\rightarrow 0.55$)}
\label{table:su2}
\newcommand{\m}{\hphantom{$-$}}
\newcommand{\cc}[1]{\multicolumn{1}{c}{#1}}
\renewcommand{\tabcolsep}{1pc} 
\renewcommand{\arraystretch}{1.2} 
\begin{tabular}{@{}c | c | c | c | c}
\hline
Decay & ${\cal A}^{(S)}$ Theory & ${\cal A}^{(S)}$ Expt & ${\cal A}^{(P)}$
Theory & 
${\cal A}^{(P)}$ Expt\\
\hline
$\Lambda\rightarrow p\pi^-$ & 1.42 (input) & $1.42\pm 0.01$ & 0.56 & $0.52\pm 0.02$ \\
$\Sigma^-\rightarrow n\pi^-$ & 1.88 (input) & $1.88\pm 0.01$ & 
$-0.50\rightarrow -0.14$ & $-0.06\pm 0.01$ \\
$\Sigma^+\rightarrow n\pi^+$ & 0.0  & $0.06\pm 0.01$ & 
$+0.42\rightarrow +0.08$  & $1.81\pm 0.01$ \\
\hline
\end{tabular}
\end{table}
At leading order, the $P-$wave amplitude in $\Lambda$ decays is well
predicted, as it is in the three-flavor theory.  However, we do not
expect significant modifications to this result from higher orders in
the two-flavor theory.  Further, we see that the $P-$wave amplitude
for $\Sigma^-\rightarrow n\pi^-$ is quite sensitive to the value of
$g_{\Sigma\Sigma}$, which is presently quite
uncertain.  At the upper limit of the allowed range for
$g_{\Sigma\Sigma}$, the $P-$wave amplitude is close to what is
observed.  Finally, we see that the $P-$wave amplitude for
$\Sigma^+\rightarrow n\pi^+$ is not close to the experimental
value for any reasonable value of $g_{\Sigma\Sigma}$, and theory
underestimates the experimental value by $\sim 4$ in the best case.

\section{NPLQCD}

The NPLQCD collaboration applied for, and was granted, time from SciDAC for
exploratory investigations of the $NN$, $NY$ and $YY$ systems ($Y$=hyperon),
and also to extract the $\sigma$-term and strong isospin breaking in the nucleon
mass.
We were awarded $\sim 8\%$ of JLab resources, $\sim 40~{\rm Gflop-yrs}$ to perform
computations with domain-wall-fermions\cite{Kaplan:1992bt,Shamir:1993zy} on the available MILC lattices,
using Chroma/QDP++ developed by Edwards and his team at JLab~\cite{Edwards:2004sx}.
We have
analyzed $I=2$ $\pi\pi$ scattering and have computed the correlation functions for
$\Lambda\Lambda$.
In what follows we present the results obtained by performing contractions with
DWF propagators generated by the LHPC collaboration, who  very kindly allowed
us to use them in our work.

\subsection{$I=2$ $\pi\pi$-Scattering}

Before tackling the two baryon systems we thought it wise to explore $I=2$
$\pi\pi$-scattering~\cite{Beane:2005rj}.
Not only is this a good ``warm-up'' exercise, but it is also interesting
physics in itself.
\vskip 0.2in
\begin{figure}[!ht]
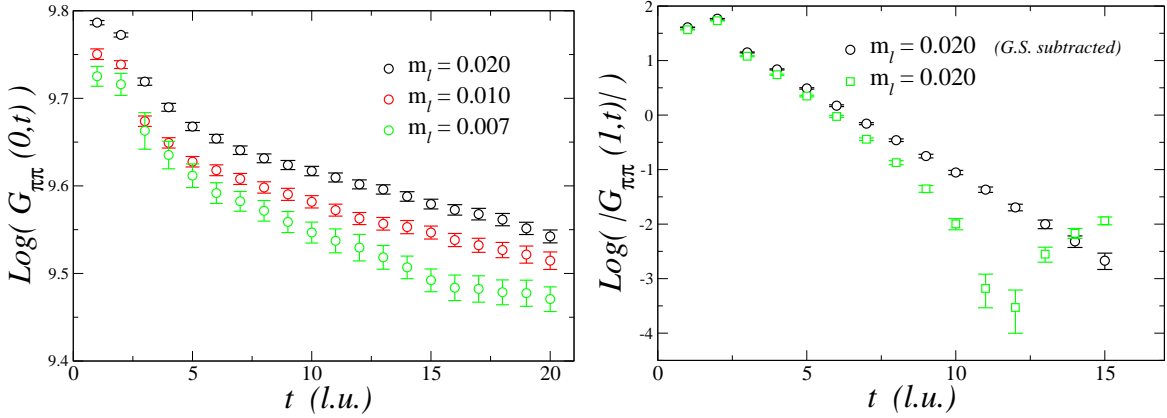

\centerline{\epsfig{file=LogpipicorrsV2.eps, angle=0,width=0.5\textwidth}
\ \ 
\epsfig{file=Logpipicorrs_excV3.eps,  angle=0,width=0.5\textwidth}}
\caption{Logarithm of the ratio of the $I=2$ $\pi\pi$ correlation function to
  the square of the $\pi$ correlation function projected onto the momentum
  $|{\bf p}|=0$ state (left panel) and the  $|{\bf p}|=1$ state (right panel).
For the excited state (right panel) we have shown the function with and without
subtracting the ground state contribution.}
\label{fig:Gpipi}
\end{figure}
The correlation functions shown in fig.~\ref{fig:Gpipi} were used to extract the ground
state and first excited state energies of $\pi^+\pi^+$ on  $20^3\times 64$ staggered
MILC lattices, with a physical dimension $\sim 2.5~{\rm fm}$ and lattice
spacing $\sim 0.125~{\rm fm}$.
The physical pion masses from these calculations are $m_\pi\sim 290, 350,$ and
$480~{\rm MeV}$.

\begin{figure}[!ht]
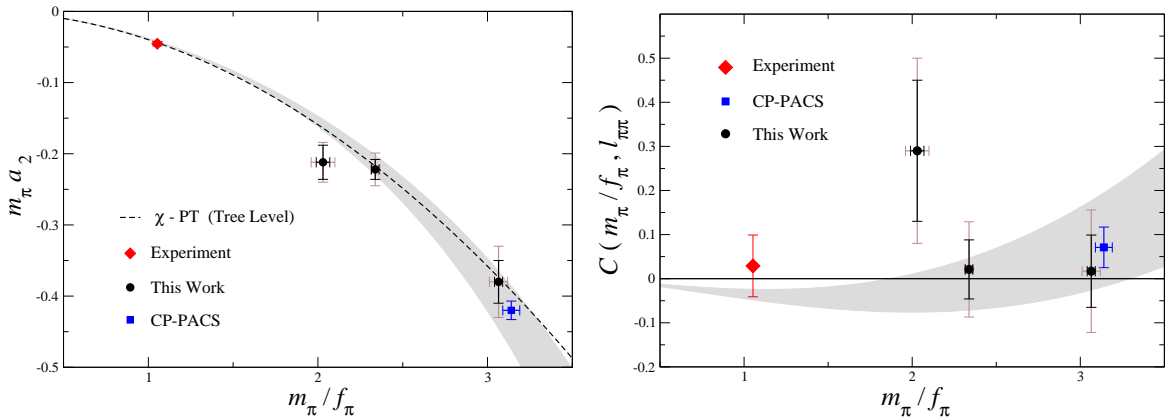

\vskip 0.3in
\centerline{\epsfig{file=mpia2PlotV4_bb1.eps, angle=0,width=0.5\textwidth}
\ \ 
\epsfig{file=CurvePlotV4.eps,  angle=0,width=0.5\textwidth}}
\caption{The results of our lattice QCD calculation 
of  $m_\pi a_2$ as a function of $m_\pi/f_\pi$
(ovals)~\protect\cite{Beane:2005rj}. 
Also shown are the experimental value 
from Ref.~\cite{Pislak:2003sv} (diamond) and the lowest quark mass result of the dynamical calculation 
of CP-PACS~\cite{Yamazaki:2004qb} (square). 
The gray band corresponds to a weighted fit to our three data points using the one-loop $\chi$-PT formula.}
\label{fig:aandC}
\end{figure}
In fig.~\ref{fig:aandC} we show the dimensionless quantity $m_\pi a_2$ versus 
the dimensionless quantity $m_\pi/f_\pi$.  The lowest mass CP-PACS point 
computed with Wilson fermions is also shown.  It is remarkable how well 
the tree-level result from chiral perturbation theory describes the lattice calculation.
Also, shown is the residual after removing the tree-level amplitude and
multiplying by $\sim f_\pi^2/m_\pi^2$.
The lattice data is not precise enough to observe the anticipated chiral logs
that enter at next order, however, at these quark masses the counterterm and
the chiral log essentially cancel.
Clearly, a more precise study of this system is required in order to identify
the chiral logs that are predicted to be present, and to allow for a
significantly more precise chiral extrapolation to compare with experiment.
Our extrapolated value is
\begin{eqnarray}
m_\pi a_2 & = & -0.0426\pm 0.0006\pm 0.0003\pm 0.0018
\ \ \ ,
\end{eqnarray}
where the first error is statistical, the second is the systematic associated
with the  data
analysis and the third is the systematic associated with the chiral extrapolation.

\subsection{$\Lambda\Lambda$-Scattering}

\begin{figure}[!ht]
\vskip 0.3in
\centerline{\epsfig{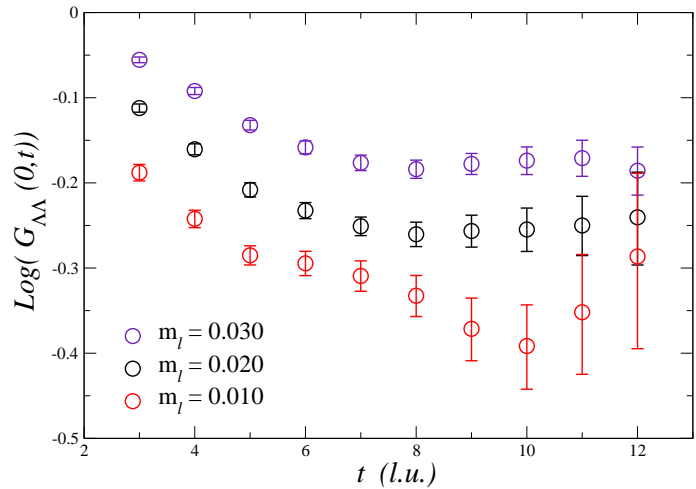}}
\caption{Preliminary data:
the logarithm of the ratio of the $\Lambda\Lambda$ correlation function to the square
  of the $\Lambda$ correlation function.
(Each has been shifted vertically by an arbitrary amount for the sake of clarity.)
}
\label{fig:lamlam}
\end{figure}

The simplest two-baryon system to write code for is $\Lambda\Lambda$, 
but this is also quite complex from a physics standpoint due to
inelastic channels that become degenerate in the SU(3) limit.
One should view this is an interpolating field for the $I=0$, $S=2$ channel.
In fig.~\ref{fig:lamlam} we show the (preliminary)
correlation functions we have obtained from some of
the MILC lattices.
The lightest quark mass correlator is quite noisy, and it will be difficult to
extract a scattering amplitude from this data cleanly.
It is clear that an increase in statistics by relatively small factors will allow for
a extraction of the scattering lengths with uncertainties much less than $\pm
1~{\rm fm}$ 
at pion masses below $\sim 500~{\rm MeV}$.

\section{Conclusions}

I am very excited by the fact that lattice QCD studies  of nuclei and
multi-hadron systems are an important part of the future of nuclear physics.
However, lattice QCD calculations alone are not sufficient, and 
it is clear that a parallel development of lattice QCD
and effective field theory efforts in this area is required for significant
progress to be made.  
Despite the large length scales that characterize nuclear interactions, the
presently available lattices allow for investigations into these
systems, and we are attempting to do this presently.

\vskip 0.2in

I would like to thank my collaborators for 
the many exciting discussions that generated this work.


\begin{thebibliography}{9}

\bibitem{Muller:1998rc}
  H.~M.~Muller and R.~Seki,
{\it Given at Caltech / INT Mini Workshop on Nuclear Physics with Effective Field Theories, Pasadena, CA, 26-27 Feb 1998}


\bibitem{Muller:1999cp}
  H.~M.~Muller, S.~E.~Koonin, R.~Seki and U.~van Kolck,
  Phys.\ Rev.\ C {\bf 61}, 044320 (2000)
  [arXiv:nucl-th/9910038].


\bibitem{Abe:2003fz}
  T.~Abe, R.~Seki and A.~N.~Kocharian,
  Phys.\ Rev.\ C {\bf 70}, 014315 (2004)
  [Erratum-ibid.\ C {\bf 71}, 059902 (2005)]
  [arXiv:nucl-th/0312125].

\bibitem{Lee:2004qd}
  D.~Lee and T.~Schafer,
  arXiv:nucl-th/0412002.


\bibitem{Maiani:ca}
L.~Maiani and M.~Testa,
{\it Phys. Lett.} {\bf B245}, 585 (1990).

\bibitem{Beane:2000fx}
  S.~R.~Beane, P.~F.~Bedaque, W.~C.~Haxton, D.~R.~Phillips and M.~J.~Savage,
  arXiv:nucl-th/0008064.

\bibitem{Haxton:1999vg}
  W.~C.~Haxton and C.~L.~Song,
  Phys.\ Rev.\ Lett.\  {\bf 84}, 5484 (2000)
  [arXiv:nucl-th/9907097].


\bibitem{Kaplan:1986yq}
  D.~B.~Kaplan and A.~E.~Nelson,
  Phys.\ Lett.\ B {\bf 175} (1986) 57.

\bibitem{Nelson:1987dg}
  A.~E.~Nelson and D.~B.~Kaplan,
  Phys.\ Lett.\ B {\bf 192}, 193 (1987).

\bibitem{Brown:1993jz}
  G.~E.~Brown and H.~Bethe,
  Astrophys.\ J.\  {\bf 423}, 659 (1994).

\bibitem{Savage:1998yh}
  M.~J.~Savage,
  Phys.\ Rev.\ C {\bf 59}, 2293 (1999)
  [arXiv:nucl-th/9811087].

\bibitem{Prezeau:2003xn}
  G.~Prezeau, M.~Ramsey-Musolf and P.~Vogel,
  Phys.\ Rev.\ D {\bf 68}, 034016 (2003)
  [arXiv:hep-ph/0303205].

\bibitem{Murphy:2000pz}
  M.~T.~Murphy {\it et al.},
  Mon.\ Not.\ Roy.\ Astron.\ Soc.\  {\bf 327}, 1208 (2001)
  [arXiv:astro-ph/0012419].



\bibitem{Rentmeester:2003mf}
  M.~C.~M.~Rentmeester, R.~G.~E.~Timmermans and J.~J.~de Swart,
  Phys.\ Rev.\ C {\bf 67}, 044001 (2003)
  [arXiv:nucl-th/0302080].

\bibitem{Kaplan:1995yg}
  D.~B.~Kaplan and M.~J.~Savage,
  Phys.\ Lett.\ B {\bf 365}, 244 (1996)
  [arXiv:hep-ph/9509371].

\bibitem{Beane:2001bc}
  S.~R.~Beane, P.~F.~Bedaque, M.~J.~Savage and U.~van Kolck,
  Nucl.\ Phys.\ A {\bf 700}, 377 (2002)
  [arXiv:nucl-th/0104030].

\bibitem{Beane:2002vs}
  S.~R.~Beane and M.~J.~Savage,
  Nucl.\ Phys.\ A {\bf 713}, 148 (2003)
  [arXiv:hep-ph/0206113].

\bibitem{Beane:2002xf}
  S.~R.~Beane and M.~J.~Savage,
  Nucl.\ Phys.\ A {\bf 717}, 91 (2003)
  [arXiv:nucl-th/0208021].

\bibitem{Epelbaum:2002gb}
  E.~Epelbaum, U.~G.~Meissner and W.~Gloeckle,
  Nucl.\ Phys.\ A {\bf 714}, 535 (2003)
  [arXiv:nucl-th/0207089].

\bibitem{Fukugita:1994ve}
  M.~Fukugita, Y.~Kuramashi, M.~Okawa, H.~Mino and A.~Ukawa,
  Phys.\ Rev.\ D {\bf 52}, 3003 (1995)
  [arXiv:hep-lat/9501024].

\bibitem{Luscher:1986pf}
M.~L{\"u}scher,
{\it Commun. Math. Phys.}  {\bf 105} 153 (1986).

\bibitem{Luscher:1990ux}
M.~L{\"u}scher,
{\it Nucl. Phys.} {\bf B354}, 531 (1991).

\bibitem{Beane:2003da}
  S.~R.~Beane, P.~F.~Bedaque, A.~Parre{\~n}o and M.~J.~Savage,
  Phys.\ Lett.\ B {\bf 585}, 106 (2004)
  [arXiv:hep-lat/0312004].

\bibitem{Bedaque:2004kc}
  P.~F.~Bedaque,
  Phys.\ Lett.\ B {\bf 593}, 82 (2004)
  [arXiv:nucl-th/0402051].

\bibitem{Sachrajda:2004mi}
  C.~T.~Sachrajda and G.~Villadoro,
  Phys.\ Lett.\ B {\bf 609}, 73 (2005)
  [arXiv:hep-lat/0411033].

\bibitem{Bedaque:2004ax}
  P.~F.~Bedaque and J.~W.~Chen,
  Phys.\ Lett.\ B {\bf 616}, 208 (2005)
  [arXiv:hep-lat/0412023].

\bibitem{Detmold:2004qn}
  W.~Detmold and M.~J.~Savage,
  Nucl.\ Phys.\ A {\bf 743}, 170 (2004)
  [arXiv:hep-lat/0403005].

\bibitem{Michael:1999nq}
  C.~Michael and P.~Pennanen  [UKQCD Collaboration],
  Phys.\ Rev.\ D {\bf 60}, 054012 (1999)
  [arXiv:hep-lat/9901007].

\bibitem{Pennanen:1999xi}
  P.~Pennanen, C.~Michael and A.~M.~Green  [UKQCD Collaboration],
  Nucl.\ Phys.\ Proc.\ Suppl.\  {\bf 83}, 200 (2000)
  [arXiv:hep-lat/9908032].

\bibitem{Cook:2002am}
  M.~S.~Cook and H.~R.~Fiebig,
  arXiv:hep-lat/0210054.

\bibitem{Cook:2005ky}
  M.~S.~Cook and H.~R.~Fiebig,
  arXiv:hep-lat/0509025.


\bibitem{Arndt:2003vx}
  D.~Arndt, S.~R.~Beane and M.~J.~Savage,
  Nucl.\ Phys.\ A {\bf 726}, 339 (2003)
  [arXiv:nucl-th/0304004].

\bibitem{Beane:2003yx}
  S.~R.~Beane, P.~F.~Bedaque, A.~Parre{\~n}o and M.~J.~Savage,
  Nucl.\ Phys.\ A {\bf 747}, 55 (2005)
  [arXiv:nucl-th/0311027].

\bibitem{Kaplan:1992bt}
  D.~B.~Kaplan,
  Phys.\ Lett.\ B {\bf 288}, 342 (1992)
  [arXiv:hep-lat/9206013].

\bibitem{Shamir:1993zy}
  Y.~Shamir,
  Nucl.\ Phys.\ B {\bf 406}, 90 (1993)
  [arXiv:hep-lat/9303005].

\bibitem{Edwards:2004sx}
  R.~G.~Edwards and B.~Joo  [SciDAC Collaboration],
  arXiv:hep-lat/0409003.

\bibitem{Beane:2005rj}
  S.~R.~Beane, P.~F.~Bedaque, K.~Orginos and M.~J.~Savage  [NPLQCD
                  Collaboration],
  arXiv:hep-lat/0506013.

\bibitem{Pislak:2003sv}
  S.~Pislak {\it et al.},
  Phys.\ Rev.\ D {\bf 67}, 072004 (2003)
  [arXiv:hep-ex/0301040].
  
\bibitem{Yamazaki:2004qb}
  T.~Yamazaki {\it et al.}  [CP-PACS Collaboration],
  Phys.\ Rev.\ D {\bf 70}, 074513 (2004)
  [arXiv:hep-lat/0402025].

\end{thebibliography}
\end{document}